# *Aligned with Whom?*
## *Direct and social goals for AI systems[1]*

Anton Korinek (Brookings, University of Virginia, and GovAI) and Avital Balwit (Oxford)

May 2022

As artificial intelligence (AI) becomes more powerful and widespread, the AI alignment problem – how to ensure that AI systems pursue the goals that we want them to pursue – has garnered growing attention. This article distinguishes two types of alignment problems depending on whose goals we consider, and analyzes the different solutions necessitated by each. The direct alignment problem considers whether an AI system accomplishes the goals of the entity operating it. In contrast, the social alignment problem considers the effects of an AI system on larger groups or on society more broadly. In particular, it also considers whether the system imposes externalities on others. Whereas solutions to the direct alignment problem center around more robust implementation, social alignment problems typically arise because of conflicts between individual and group-level goals, elevating the importance of AI governance to mediate such conflicts. Addressing the social alignment problem requires both enforcing existing norms on their developers and operators and designing new norms that apply directly to AI systems.

**Keywords:** agency theory, delegation, direct alignment, social alignment, AI governance

# I. Introduction

We are building artificially intelligent systems which are increasingly competent and general. Already, AI systems can have impacts that occur almost instantly and anywhere in the world.[2] While speed and reach are factors that AI shares with other technologies like malware or nuclear weapons, AI systems are also different, becoming increasingly autonomous and behaving like agents of their own.

When we delegate tasks to AI systems, an important challenge is to endow the systems with "desirable goals" — this is frequently labeled the AI alignment problem. For a good survey of recent research on AI alignment see, e.g., Ngo (2020). However, when speaking of alignment, it

---

[1] This paper was prepared for the *Oxford Handbook of AI Governance*. We would like to thank Ondřej Bajgar, Damon Binder, Justin Bullock, Alexis Carlier, Carla Zoe Cremer, Allan Dafoe, Ben Garfinkel, Lewis Hammond, Fin Moorhouse, Luca Righetti, Toby Shevlane, Joseph Stiglitz, and participants at the Spring 2021 Handbook of AI Governance conference for helpful comments and discussions. Any remaining misalignment is our own. Korinek acknowledges financial support from the David M. Rubenstein Fellowship program at the Brookings Institution. The views expressed herein are those of the authors and do not necessarily reflect the views of the Brookings Institution.
[2] An example are AI-powered financial trading systems (see e.g. Boukherouaa and Shabsigh, 2021).

is important to be explicit about whose goals an AI system is aligned with. This paper distinguishes between two types of AI alignment that are both important in AI development but that require distinct governance approaches:

**Direct alignment:** when an AI system is pursuing goals consistent with the goals of its operator, irrespective of whether it imposes externalities on other parties.[3]

**Social alignment:** when an AI system is pursuing goals that are consistent with the broader goals of society, taking into account the welfare of everybody who is impacted by the system.

There are multiple challenges to achieving direct alignment, including identifying the right goal to give to an AI system, conveying that goal, and getting the AI system to correctly implement the goal.[4] Social alignment adds the challenge of including all of those impacted — not merely the operator of a system — to the extent that society views this as desirable when determining what goal to give the system.

What would it entail to consider all those affected? We describe a conceptual benchmark for social alignment based on welfare economics that we call *ideal social alignment* and analyze how social alignment differs from direct alignment in that it considers externalities. We provide examples of ideal social welfare functions that could, in principle, be employed to formally study social alignment. Since ideal social alignment is unattainable in practice, we explain how it is possible to determine a partial ordering of social preferences and that in practice, social alignment means aligning AI systems with this more limited set of instructions. Social alignment will generally require external interventions and a broader governance framework. We discuss how regulatory solutions provided by social norms, laws, markets, and architecture could be used to achieve social alignment, and we highlight the importance of AI governance to achieve social alignment.

We believe ensuring that AI systems satisfy both direct and social alignment will become ever more important in the future. Currently, we delegate fairly narrow tasks to AI systems — we ask them to classify images, rank our search results, recommend movies and music, and autocomplete bits of our emails. But we are starting to use AI in increasingly higher stakes environments, like evaluating loan and job applicants, financial trading, utility management, and national defense. Importantly, the range and complexity of tasks that we turn over to our AI systems are continuing to grow.

AI Alignment — the challenge of how to endow AI systems with goals that are consistent with our goals — becomes ever more important as delegation from humans to AI systems (i) happens more often, creating more opportunities for misalignment to cause harm; (ii) involves higher stakes situations, where misalignment would be more costly or even catastrophic; and (iii) occurs in situations where we have less ability to provide oversight, which makes it more difficult to assess whether the system is aligned and less likely that we catch alignment failures early.

---

[3] Throughout the paper, we use the convention of referring to the entity that is creating, operating, and controlling an AI system as the "operator." In principle, each of these tasks could be performed by different entities, adding additional complexity to the challenge of AI alignment.

[4] There are many alternative ways of defining alignment but with similar flavor. For example, an AI system could be aligned to the human's instructions, intentions, revealed preferences, informed preferences, interests, or values, among other options. See Gabriel (2020) for a fuller discussion.

## II.  Concepts

We start by clarifying several theoretical concepts that are relevant for our discussion of alignment. Readers who are most interested in the comparison of direct and social alignment may wish to skip to the following section.

### Agents and goals

**Agents:** We call entities that interact with their environment "agents" whenever it is useful to describe them as pursuing goals.[5] A goal is a summary description of what an entity is attempting to achieve. Humans clearly fit our description of agents—we think of ourselves and each other as pursuing goals, and this is useful because inferring people's goals makes for a more efficient description of what to expect from them next. For example, if a driver sees a person walking straight up to a crosswalk, it is useful to infer that the person's goal is to cross the street, and this is more efficient than to ponder how the person's leg movements will translate into the person's physical location over the ensuing seconds.

But our concept of agency is broader than humans. It also includes non-human entities such as organizations or governments, which can be described as pursuing their own sets of goals. For example, a business organization may be described as following the goal of producing a product to earn profits; a university as advancing research and education; a government as pursuing the safety and well-being of its citizens.

And, importantly for the purposes of this article, our concept also includes artificial intelligence. Many modern AI systems are directly programmed to maximize a specific objective function, making them act in a goal-oriented way. More broadly, AI systems are agents in the sense that we define because they are designed to pursue specific goals, for example classifying images, driving cars, controlling robots etc.

In our characterization of agency, we explicitly ask whether it is "useful" to describe an entity as pursuing a goal. This implies that the delineation is fluid and depends on the context. For example, if we start with a very simple mechanical structure and transform it into a progressively more intelligent robot, there is no specific threshold at which it becomes an agent—but it will become more and more efficient and useful to describe it according to the goals it pursues rather than by the physical laws describing its mechanical structure. To provide further examples, it will rarely be useful to describe a rock as an agent, but there will be many situations in which it is useful to describe a Boston Dynamics Spot robot[6] as an agent. Moreover, the same entity may be usefully described as an agent in some situations and contexts but not in others. For example, we may want to describe a robot as an agent while it is operating but as a piece of metal when we recycle it for scrap.

---

[5] This is a shallow definition of agency that is, however, useful for our purposes here. It is inspired by but distinct from Dennett's work on stances (see e.g. Dennett, 1987). In different contexts, other definitions may be more useful - for example, in ethics, a moral agent is an entity that is morally accountable for its actions. For an elaboration on alternative concepts of agency, see e.g. Franklin and Graesser (1996) or Orseau et al. (2018).

**Goals:** In the social sciences, an entity's goals (or objectives) are frequently described using a set of preferences, i.e., an ordering that describes how the entity values different outcomes relative to each other. For example, a preference relation such as A > B reflects that the entity prefers outcome A over outcome B.[7]

There is a dualism between an agent's goals and her actions—we can either describe the actions which emanate from her goals or describe the goals which drive her actions. When an agent's goals are fully specified in a given environment, the agent's actions are also fully specified (except where there are ties) because we can work out what actions the agent will find optimal to take, and vice versa because we can infer what their goals are from their actions. We can simply translate back and forth from goal space into action space.

Simple depictions of agents in the social sciences frequently take an agent's goals as a primitive that is exogenously given, for example by postulating a certain utility function for the agent. Then they proceed to analyze what actions the agent would take to achieve the most desired outcome. However, such descriptions risk over-simplifying the behavior of human agents by leaving out subgoals, competing goals, or constraints on the described goal. For example, humans track the distance between current and desired states across multiple value dimensions that are in tension with each other in ways that change dynamically over time (see e.g., Juechems and Summerfield, 2019). This may pose significant challenges in determining human goals and may, for example, produce behavior that is inconsistent with any utility function.[8] As we will discuss in more detail below, the difficulty of correctly determining goals is a major challenge for AI alignment.

---

[6] See https://www.bostondynamics.com/products/spot
[7] Another way to express goals is in the form of a "utility function" u(X) that assigns a numerical value to each possible X which ranks the different possibilities. Utility functions are more restrictive than preference orderings. In other words, every utility function defines a set of preferences, but not every set of preferences can be captured by a utility function. For example, for lexicographic preferences, doing so is impossible.
[8] Specifically, for a set of preferences to map into a utility function requires several technical assumptions that may be violated, including completeness, transitivity, and continuity.

# Delegation and alignment

Delegation is when an entity charges another with the fulfillment of her goals. Following the language of economics, we use the term "principal" for the entity that delegates a task and "agent" for the entity who is charged with the task.[9]

Examples of delegation are as old as humanity itself and could already be found in hunter gatherer societies.[10] As societies became more complex and hierarchical, the delegation of tasks and the need for goal alignment also started to involve governmental or religious organizations and, later, corporations. Humans and organizations both started to delegate tasks to each other to better accomplish their goals. For example, entrepreneurs founded corporations to pursue their goals, and corporations hired workers.

Successful delegation is advantageous for the principal by enabling her to better accomplish her goals. The reasons for this advantage include that the agent may have a comparative advantage in the task at hand, such as different or greater capabilities than the principal, better knowledge, or simply a lower opportunity cost of time. To make delegation successful, there needs to be a sufficient degree of alignment between the principal and the agent, but the principal also needs to provide the agent with a sufficient degree of discretion—a point that is not typically emphasized in traditional principal-agent theory in economics but that was already made by Weber (1922) and that is also emphasized in the recent literature on AI governance (see e.g., Young et al, 2021). In particular, the agent needs sufficient freedom of action to employ her greater capabilities, her better knowledge and judgment or her additional time to be useful to the principal. However, this freedom of action is what creates problems when there is misalignment.

Economists and other social scientists have long studied principal-agent relationships, i.e., how to align the actions of agents with the outcomes desired by principals (see e.g. Jensen and Meckling, 1976, for one of the most influential contributions). In that body of work, the principal and agent have exogenously given goals that differ from each other, and imperfect information makes it difficult for the principal to observe whether the agent has acted in her best interest or has abused the discretion she was afforded. The main research question centers around how the principal can provide the agent with incentives to act in her interest—for example, how to include carrots and sticks in work contracts to incentivize workers to exert the optimal level of effort.

---

[9] Unfortunately this nomenclature involves some overloading of the term "agent" - in the previous section, we called any entity that can be described as pursuing a goal as an agent; in this section, we follow the conventions of principal-agent theory. Throughout the remainder of this article, the meaning of the term will be clear from the context.

[10] In fact, some of the interactions among other social species such as bees or ants can also be described as simple forms of delegation.

In AI alignment, by contrast, the question is different and in some sense more fundamental: how to endow an agent with goals that lead to outcomes that are desired by the principal.[11] AI alignment is usually described as goal alignment, but what ultimately matters for the principal are the agent's actions. When the relationship between goals and actions is clear and is known perfectly and when the goals of principal and agent coincide, then there is no direct alignment problem. In Figure 1, this would correspond to each of the mappings that are indicated with arrows holding perfectly, i.e., (1) the principal has identified the goals that will lead to her desired actions well, (2) they are correctly conveyed to the agent, who (3) in turn translates the conveyed goals into the desired actions. However, this is an idealized benchmark.

| Delegation | Principal | | Agent |
|---|---|---|---|
| Action space | Desired actions | | Pursued actions |
| | $\updownarrow$ (1) | | $\updownarrow$ (3) |
| Goal space | Desired goals | $\leftrightarrow$ (2) | Pursued goals |

Figure 1: Principal-agent alignment in action space and in goal space

In practice, misalignment between the principal's desired and realized outcomes can arise in any of the three steps outlined in the figure. This allows us to distinguish the source of alignment problems into the following categories:

(1) **Identifying the principal's desired goals:** The principal needs to figure out what her goals are.
(2) **Conveying the goals to the agent:** Next, the principal needs to correctly transmit her desired goals to the agent so the agent can pursue them.
(3) **Translating the goals into actions:** The agent needs to correctly implement the transmitted goals by pursuing the corresponding actions.

---

[11] On the surface, the two described situations - incentivizing a human agent with distinct goals to pursue the principal's goals versus creating an AI agent from scratch who pursues the principal's goals - seem very different. However, given the dualism between actions and goals, there is in fact a deeper equivalence between the two. Addressing the classic principal-agent problem in economics can be viewed as a situation in which the principal has only limited ability to affect the agent's architecture (e.g. to reprogram the agent's primal drive to avoid hard work) and needs to find workarounds ("incentives") to make the agent pursue the desired goal. Programmers frequently experience similar situations. For example, the architecture of ML libraries, say TensorFlow, constrains how they can write their code and makes some results far easier to obtain than others. In other situations, they need to write workarounds building on clunky legacy applications to efficiently obtain the desired behavior.
Conversely, human principals sometimes have the ability to "reprogram" agents. For example, parents greatly appreciate the importance of instilling proper goals into their offspring; managers and military leaders know the importance of "inspiring" their agents to pursue desired goals; and a significant part of our human culture (religion, morals, etc) revolves around reprogramming humans' goals in a way to make our societies operate more harmoniously.

Below, we will elaborate on each of these steps in detail.

As we observed before, the term "AI alignment problem" is frequently used to describe step (2), or a combination of steps (1) and (2), i.e., how to provide an AI system with a set of goals that correctly reflect our goals, whereas the term "AI control problem" is used to capture the broader challenge of how to ensure that the actions of an AI system are desirable.[12] However, the three steps illustrated in our figure are closely related to each other. If the mappings described in steps (1), (2) and (3) held perfectly, it would be possible to focus exclusively on how to align the goals of human principals and AI agents. In practice, however, it will be necessary to consider all four steps simultaneously as they cannot be cleanly separated from each other.

## Direct and social alignment of AI

So far we have discussed how to ensure that the actions performed by an AI system reflect the desires of a principal. However, we have been silent on who exactly the principal is. We distinguish two separate concepts, direct and social alignment, that relate to whether we view the principal as the operator of an AI system or as society at large. The distinction between the two has not been sufficiently recognized in the existing literature on AI alignment.

As defined in the introduction, we use the term **direct alignment** to refer to whether an AI system is pursuing goals that are consistent with the goals of its operator, and **social alignment** to refer to whether an AI system is pursuing goals that are consistent with the broader goals of society, taking into account everybody who is affected by the system and internalizing any externalities. The two forms of alignment problems are also related to the broader challenge of developing cooperative AI (see e.g., Dafoe et al, 2020).

In the following two sections, we will discuss the challenges of direct and social alignment in detail and will elaborate further on when the two differ from each other. But before doing so, let us provide two examples to highlight the difference between the two.

*Example 1:* Ted develops a new resume screening algorithm. Since he wants the system to be free of racial bias, he leaves out the variable "race" from the training dataset. However, the system quickly learns correlates of race such as name, address, and educational institution from the existing bias in the training dataset and uses these to arrive at biased hiring decisions.

This represents a failure of direct alignment. Ted was eager to avoid racial bias but did not realize that his implementation led to precisely the bias that he was concerned about.

*Example 2:* Mark develops a recommendation model to maximize user engagement on a social network platform. When he finds out that the system leads to stark increases in political polarization, he does not change course.

---

[12] For example, Bostrom (2014) describes value alignment as one element of AI control alongside other mechanisms such as capability controls.

This represents a failure of social alignment. The AI system pursued—and successfully achieved—its assigned goal, but it imposed large externalities on society by increasing polarization.

# III. Direct alignment

From the perspective of a principal operating an AI system, there are three interrelated challenges to ensuring direct alignment of an AI system: the first is the challenge of determining what goals to pursue, the second is conveying the goal to an AI system, and the third is getting the AI system to correctly translate the goal into actions.

## Determining the goal

The first challenge is to work out the principal's goals, i.e., to translate the principal's desired outcomes into abstract goals. Determining what we want can be difficult. Human goals are not easily interpretable; they are often amorphous or intuitively understood but difficult to express. Our brains are opaque and pursue multiple value dimensions depending on circumstances (see e.g., Juechems and Summerfield, 2019). When we are not sure what we want, it is difficult to align an AI system with our goals.

A key aspect of determining the principal's goal is how to scope the goal appropriately so it does not conflict with other goals that are valuable to the principal. Humans have a broad set of goals that involve many different subgoals that we automatically and often subconsciously weigh against each other when they are in conflict. When we determine what goals to assign to AI systems, we have to ensure that the systems do not optimize one subgoal to the detriment of others. This becomes more and more important as our AI systems become more powerful and their capacity to optimize over a single goal increases.

*Example 3:* Mark develops a recommendation model to maximize user engagement on a social network platform. He is dismayed to find out that the system also increases political polarization.

This is a classic example of specifying an excessively narrow goal and obtaining unexpected side effects. In the described example, Mark did not anticipate that his recommendation model would also affect the political views of its users.

## Conveying the goal

After a principal determines the content of her goals, she faces the technical challenge of transmitting the goal to an AI system. When humans convey goals to each other in natural language, they understand the context, which makes it easier to resolve ambiguities.

It is more challenging to convey goals to AI systems. AI systems do not share the same understanding of the world that we humans share and will likely not be able to resolve ambiguities by making common-sense deductions. Instead, we need to translate the goal into something machine readable and provide the instructions, training, or feedback necessary for the system to "understand" and execute. Part of the challenge is to clarify what our concepts mean. For example, if we told an AI system to "make us happy," what do we mean by that term? Do we mean pure hedonic experience, do we mean general life satisfaction, or any other range of viable meanings?

*Example 4:* Tim tells his AI-powered smartphone assistant to "call Jon" as he gets ready to go out and party. He is embarrassed that his 11 p.m. phone call wakes up his boss rather than reaching his brother, who is listed as "Jonathan" in his contact list.

This is an example of an AI system misinterpreting the goal of its principal because it did not correctly understand the context. A human assistant would have known not to call a work contact late at night and would have understood that "Jon" may refer to "Jonathan."

Sometimes, the challenges of determining and conveying can be addressed jointly. For example, inverse reinforcement learning allows an AI system to learn the objective function of its principal through observing their behavior (Ng and Russell, 2000). Through this method, the content and form of the goal blend together.

## Implementing the goal

Once the principal's goals are transmitted to the agent, they need to be implemented through appropriately chosen actions by the AI system. Implementation has many technical aspects that we will not discuss here.[13] These include ensuring a system is free of bugs and robust, including that it operates reliably in circumstances other than what it was initially trained and tested on. While implementation is key for getting a result that the principal is happy with, some do not view it as a pure component of alignment.[14]

Example 1 above described a resume screening system that resulted in biased hiring decisions because of biased training data that was not corrected for. This is an example of an implementation failure.

## Distinguishing direct alignment from social alignment

The described challenges of direct alignment also apply to social alignment. The key difference is that the first step, determining the goal, no longer involves a single principal who is operating the system, but instead the broader goals of others in society who would be impacted by the AI system. This is what makes the social alignment of AI a central theme of AI governance.

---

[13] For a thorough and cutting-edge technical introduction see e.g. Russell and Norvig (2020).

[14] For example, some definitions of alignment only capture the intentions of the AI system, not the outcome, whether ex ante the AI system was trying to achieve the human's goal. If an AI system tries to accomplish the goal, but some implementation failure causes the system to crash before doing so, perhaps it should not be viewed as a failure of alignment.

Many contributions in the existing literature on AI alignment refer to either direct alignment or social alignment without explicitly addressing the distinction between the two and the necessity of paying attention to both. For example, Paul Christiano (2018b) appears to focus largely on direct alignment in his definition of intent alignment: "AI A is aligned with an operator H, if A is trying to do what H wants it to do." It is possible that the operator in this definition wants the AI to do something that will impose large externalities on others. Yudkowsky (2004) seems to describe a form of social alignment in defining an alignment benchmarked that he terms coherent extrapolated volition (CEV) as "our wish if we knew more, thought faster, were more the people we wished we were, had grown up farther together..." He explicitly refers to "*our* wish...had [*we*] grown up farther *together*."

Other definitions are unclear on whether they imply direct or social alignment. Christiano (2018a) defines the project of alignment as "aligning AI systems with human interests," leaving unspecified which humans these interests belong to. For example, Stuart Russell (2019) defines the problem of value alignment as ensuring that we do not "perhaps inadvertently, imbue machines with objectives that are imperfectly aligned with our own" (p. 149). Brian Christian (2020), author of "The Alignment Problem," defines the titular problem as "how to ensure that these models capture our norms and values, understand what we mean or intend, and, above all, do what we want" (p. 19). Both of these definitions do not make it clear who "we" or "our" refers to—all of humanity, some representative group, a synonym for one? Evan Hubinger (2020) writes that an AI agent is "impact aligned (with humans) if it doesn't take actions that we would judge to be bad/problematic/dangerous/catastrophic." It is not clear who these "humans" are—just the principals/operators, or all humans—or how large a consensus this "we" has to attain.

# IV. Social alignment

In contrast to the direct alignment problem, which looks at whether an AI system is consistent with the goals of its operator, the social alignment problem looks at whether the system pursues goals that are consistent with the goals of society at large.

In the following, we start by describing a benchmark for social alignment based on welfare economics that we call *ideal social alignment*. We analyze how social alignment differs from direct alignment by emphasizing that it considers externalities on others. We provide examples of ideal social welfare functions that could, in principle, be employed to formally study social alignment. However, we also lay out the practical difficulties of establishing what society's preferences are, and we analyze how to deal with situations in which there is no clear way of mediating conflicting objectives within society. See also Baum (2017) for a description of the difficulties in establishing society-wide ethics for AI.

Next, we analyze to what extent existing norms (broadly defined) for the operators of AI systems are sufficient to ensure that the systems are socially aligned. We distinguish two sets of problems that give rise to misalignment: The first arises when the operators of AI systems violate existing social norms and are themselves not socially aligned; the second arises when our current social norms are insufficient to address some novel externality caused by new AI capabilities. Achieving social AI alignment requires progress on both fronts. We describe a set of policy tools that can be used to further the social alignment of AI.

## Ideal social alignment

To describe a simple idealized benchmark for social AI alignment, we impose two assumptions. First, we assume that society has a complete and well-defined set of preferences over all choices that are relevant for the AI system. Second, we assume that society itself creates, operates, and fully controls the AI system. Under these idealized conditions, the challenges of social alignment of AI would be reduced to the challenges of direct alignment that we described in the previous section, since society would essentially act as the operator.

Technically speaking, the first assumption of complete preferences implies that society's preferences define a full ordering of all the available choices, i.e., that whenever society faces a set of choices, say A, B or C, there are social preferences that tell us how they rank compared to each other, which ones are preferred, or which ones are equally desirable for society.[15] One possible ranking would be for example A > B > C. As we will explore further below, this is a strong assumption; in practice, there are many areas in which society does not have such clear preferences.

Theoretically, ideal social alignment could also be achieved if an AI system is operated by a true altruist who implements society's preferences. Most human beings are not purely egoistic but also care about the well-being of others—they have altruistic preferences, meaning they explicitly consider and accommodate the preferences of a broader group than just themselves. Their consideration of the welfare of others may range from minor, i.e., they care about others much less than about themselves, to perfect altruism, i.e., their decision making considers the welfare of another person just as much as their own. The direct alignment problem only considers the effects of the AI system on others to the extent that the person or entity creating the system is altruistic. If those who set AI goals are perfectly altruistic, like this theoretical altruist, then the social alignment problem is solved—their individual goals will be aligned to social goals. However, whenever the preferences that the altruist perceives about society differ from society's actual preferences—or whenever society's preferences change—the supposed altruist would become a dictator. This is a significant risk whenever there is no democratic process that represents a corrective force to express society's preferences.

---

[15] In mathematics, a full ordering (or total order) is a binary relation on a set that satisfies, among other conditions, that it is transitive and that any two elements are comparable. See e.g. https://en.wikipedia.org/wiki/Total_order The assumption that society's preferences represent a full ordering is also a necessary condition for describing them via social welfare functions.

As reflected in the term, *ideal social alignment* is more a conceptual benchmark than a guideline for actual implementation. Even in a well-functioning representative democracy, it is unlikely that society will agree on all outcomes. However, the benchmark is useful to sharply delineate the differences between direct and social alignment.

**Externalities:** At the center of the distinction between direct and social alignment are externalities. In our context, externalities exist whenever an AI system affects others without their agreement and without the beneficiary compensating others for it. For example, an AI system that engages in surveillance of individuals or otherwise intrudes into their privacy imposes externalities onto them. Likewise, a system that manipulates consumers into buying goods on false pretexts inflicts externalities on its victims. Similarly, a system that evaluates applicants for jobs, loans, or other transactions but engages in unfair discrimination imposes externalities on its subject. At a society-wide level, a system that promises social engagement to its users but creates echo chambers and deteriorates democratic processes while doing so inflicts significant externalities on society at large. Similarly, AI systems that displace human jobs without generating sufficient new jobs depress economy-wide labor demand, reduce wages, and thereby contribute to rising inequality. These effects represent what economists call pecuniary externalities because they arise from changes in market prices (in the given case, declines in workers' wages). AI systems may also confer positive externalities on society—for example, a tool that aids in scientific discovery may have benefits for society that are far greater than the benefits that the developer of the tool obtains.

Framed this way, the goal of social alignment is to internalize the externalities of AI systems—making sure that AI systems consider the benefits and costs not only for their operators but also for all other members of society. In some cases, society may view the potential externalities of an AI system as so harmful that it is best for the system not to be implemented. For example, many forms of mass surveillance may simply be too harmful to be worth any benefits they create. In general, however, it is not necessarily desirable to reduce all externalities to zero, just to properly take them into account. For example, an AI system that creates large amounts of value at the cost of displacing some jobs may well be worth implementing. Social alignment simply requires that the balance of the useful and deleterious effects reflects what society would choose as the optimal balance.

**Social welfare functions** are a widely used concept to weigh benefits and costs that can be applied for the ideal social alignment of AI systems. Social welfare functions are a generalization of utility functions to social choices and work in a similar fashion: Given a set of possible choices for society, a social welfare function assigns to each choice a welfare score that captures how desirable it is for society, and the choice that obtains the highest score is by definition the preferred one that society would adopt. A widely used example are utilitarian (also called Benthamite) welfare functions, which sum up the utilities of all members of society with equal weight. A specific case among utilitarian social welfare functions that evaluate consumption choices is when individuals are attributed utility functions that are linear in consumption—in that case, social welfare is simply the sum of all consumption, and maximizing social welfare is equivalent to maximizing economic efficiency, i.e., maximizing total consumption without any regard for distributive concerns. Conversely a Rawlsian (or maximin) social welfare function captures the desire to maximize the utility of the worst-off member of society, reflecting a strong desire for equality.

The analogy to utility functions makes social welfare functions an appealing concept for AI developers—utility maximization is a cornerstone of training AI systems. This makes it important to be aware of the limitations of social welfare functions. One limitation of social welfare functions as commonly used is that they focus on consequentialist specifications of individual utility, which may miss other considerations that society views as important for ethical decision making, e.g., deontological considerations.[16] Imposing a bad welfare function, even with the best of intentions, will create social harm. Another important limitation is that social welfare functions pre-suppose a complete set of social preferences that are well-defined over all choices, but there are many situations in which this may not be achievable. This is what we will turn to next.

## Disagreement and partial social preferences

Social alignment can only occur where there are well-defined social preferences. However, members of society frequently disagree on what are the most desirable choices. Unfortunately, there is no general way to compile the preferences of multiple individuals into well-defined and rational social preferences over all available choices. Condorcet (1785) observed that democratic voting will not in general produce a full set of rational social preferences.[17] Arrow (1950) showed in his doctoral thesis that this is a general property of all mechanisms to aggregate individual preferences into social preferences, except for a dictatorial rule whereby a single member of society dictates all choices.[18] These negative results imply that it is generally not possible to come up with the complete set of preferences that would be necessary for AI systems to implement the ideal social alignment that we described above. However, that does not imply that we need to give up on social alignment entirely.

Instead, social alignment can still focus on aligning AI systems with those social preferences that can be clearly established. Even though members of society may not agree on how to rank all available choices, they will agree on how to rank many of the most important choices. They may agree that A > B and A > C but may not be able to rank B and C relative to each other. The social choices over which there is general agreement within society represent a partial ordering of all the available choices. We can increase the set of choices for which a social preference can be established if we weaken the standard from universal agreement to somewhat lower standards, e.g., near-universal agreement. For example, society will generally agree that it is desirable to save lives or to refrain from actively discriminating against minorities, even if a small fraction of the population disagrees. The resulting partial ordering provides a limited set of instructions for social choices—even though it cannot identify a full set of social preferences that apply to all circumstances.

---

[16] Although it is possible to add such considerations with a negative weight in consequentialist specifications of welfare functions, it is difficult to determine desirable weights.

[17] Specifically, when three or more people are asked to express their preferences over three or more alternative choices in pairwise votes, they frequently arrive at outcomes like A > B, B > C and C > A, making it impossible to establish a full order of the available social choices.

[18] Even seemingly straightforward mechanisms such as a welfare function that is the sum of utility functions of all members of society will not satisfactorily address the problem, since it may lead to Pareto-dominated outcomes. See Eckersley (2019) for a fuller description of the problem in the context of AI alignment.

Social alignment requires that AI systems observe the partial ordering provided by social preferences. Formally, we call an AI system socially aligned if its choices correspond to the partial ordering implied by social preferences. Conversely, an AI system violates social alignment if it makes choices that contradict the partial ordering implied by social preferences, i.e., if society generally agrees that it would make different choices.

Given the partial nature of the ordering implied by social preferences, there will be situations in which society genuinely disagrees, so social preferences do not provide instructions for what a socially aligned AI system should do. Existing social norms cannot provide guidance for the AI system or its operator what choices to make. In some contexts, there may not be a preferred social choice, but there may be agreement among members of society that AI systems and their operators should have the liberty and freedom to make their own choices, for example the freedom of how to design a new product or how to compete in the market (while respecting the rules). In open societies, such freedoms are in themselves an important value.

However, in other contexts, unresolved conflicts within society imply that the choices of AI systems and their operators are likely to be contentious, no matter what choice they make. One of the fundamental goals of governance is to resolve such conflicts, to determine how to establish social preferences in such situations, and how to establish norms that encapsulate these preferences. The question of what social preferences ought to be in the realm of AI—and over which social choices they are defined—is a central theme of AI governance.

Rights-based approaches are a common mechanism by which society represents partial social preferences because they reflect certain defined entitlements and freedoms for members of society while leaving ample space for other choices over which there may be disagreement. For example, one of the most fundamental rights-based approaches, human rights, encapsulates a set of basic rights and freedoms for which there is general agreement among the countries adopting the declaration that all human beings are entitled to them (Universal Declaration of Human Rights, 1948). In the digital realm, an example is the EU's General Data Protection Regulation (GDPR, 2016), which adds many new rights that have become relevant only recently, for example the right of access to information or the right to be forgotten. Bajgar and Horenovsky (2021) describe how rights-based approaches may be useful for long-term AI safety and regulation.

## Spheres of social alignment

The social alignment problem can manifest at several different scales, ranging from small subgroups of society to larger spheres such as humanity as a whole. The commonality between them is that they all include cases where someone other than the AI system's operator is affected by externalities.

In general, the extent of agreement within a group decreases with group size. Smaller communities may find it easier to come to an agreement on what outcomes are desirable for an AI system to pursue than larger groups of people such as the citizens of a nation or humanity as a whole. This implies that group preferences will lead to increasingly partial orderings as the group size increases, i.e., larger groups will agree less on what outcomes to pursue than smaller, more homogenous groups.

A good example of how attitudes towards social alignment differ depending on group size is when there are competitive dynamics between subgroups of society. Consider two large corporations developing AI systems that are in fierce competition with each other—when one corporation improves its system, it gains market share at the expense of the other. The stakeholders of each corporation, including its workers, shareholders, and suppliers, form a subgroup of society, and this subgroup has a clear interest in the corporation doing well. If a corporation's AI system pursues outcomes that are in the collective interest of that subgroup, then the system is socially aligned at the subgroup level, i.e., at the level of the corporation.

Looking outside the corporation, there are clearly negative externalities between the two corporations. However, our social norms at the national and international level allow for such competition and do not find competitive dynamics objectionable as long as they benefit consumers and satisfy other applicable laws. The partial ordering reflecting our society-wide norms includes the requirement to act lawfully, to avoid biases, etc., but gives corporations freedom to engage in lots of actions, including the freedom to compete with each other. Under the described circumstances, the corporation's AI system is aligned at the society-wide level even though it imposes large negative externalities on competitors.

For purposes of illustration, and without being exhaustive, we discuss a few different exemplary spheres of social alignment:

**Group-level:** A social alignment problem could manifest at the group level—for example, within a community, club, university, corporation, or city. If an AI system run by members of the group imposes externalities on the group that violate the social norms within the group, it is socially misaligned at the group level.

**Country-Level:** For many questions, individual countries are the most important sphere at which to consider social alignment. In the modern world (perhaps with the exception of the EU), most laws and regulations originate at the country level since countries are the politically most powerful actors. This is also true for AI regulation to forestall alignment problems.

However, social alignment at the country level is not necessarily sufficient—countries are frequently subject to competitive dynamics, especially in the military context, where advances in AI may give rise to significant shifts in power dynamics (see e.g., Armstrong et al, 2016). This directly leads to the next and widest sphere at which social alignment is desirable.

**World-level:** Social alignment at the world level is the broadest, least restrictive, but perhaps also most fundamental sphere of alignment for AI systems. It requires that an AI system pursues desired outcomes on which humanity at large broadly agrees. Although there are many areas of significant disagreements among the world's citizens, there are also areas of almost-universal agreement. Examples include the desirability of basic forms of AI safety to avoid human extinction, or that the most momentous decisions undertaken by autonomous weapons systems should have humans in the loop (e.g. Human Rights Watch, 2012). Another area of near-universal agreement may be that it is undesirable to develop a super-human AI system that displaces all human labor without ensuring that humans have sufficient material resources to survive such a radical shift. Articulating and formalizing global social norms on these topics is a pressing area of concern.

## Implementing social alignment

No matter in which sphere, attaining social alignment of an AI system may be more challenging than attaining direct alignment—the main goal for the creator of an AI system is to solve the direct alignment problem so that the system pursues the outcomes he desires. Social alignment may be an afterthought.

The social alignment of an AI system would be ensured if its operator is perfectly altruistic and internalizes all externalities that the system imposes on others in an ethical fashion. However, more generally, the operator may not care about imposing harm on others as long as the system achieves the outcomes they desire for themselves. In the following, we discuss the available avenues to achieve social alignment.

**Assessing AI impacts:** A precondition for evaluating social alignment is knowing about the impacts and potential externalities generated by an AI system. Sometimes harms arise without the operator of an AI system even being aware of it, and the operator may not have sufficient incentives to find out. Moreover, lack of transparency makes it easier to cover up harms. AI impact assessments could help. Lessons can be learned from environmental impact assessments (EIAs), which are routinely required for actions of government agencies, or for government-funded, -permitted or -licensed activities—for example, for building a highway, airport, or oil pipeline. AI impact assessments could quantify the potential risks and benefits of AI systems. Such assessments could be mandated for AI projects that are implemented by government entities, that receive government funding, or that have sufficiently broad societal effects. They could also be used on a voluntary basis, just like EIAs have become relatively common in the private sector.

**Existing norms:** Once the potential externalities of an AI system are known, the next question is how social norms, regulations, and laws can ensure the social alignment of AI systems. Society already has a rich set of norms for social alignment that have evolved over centuries, consisting of informal social customs and habits as well as formal laws and regulations. These norms constrain the behaviors of both individual humans and non-human agents such as governments, corporations, or nonprofits. Whenever these entities operate AI systems, the existing norms that they are subject to by extension represent norms for the behaviors of their AI systems. To provide a stark example, a civilian must not program an AI-based robot to kill someone.

The social alignment of the operators of AI systems thus leads to a certain "default" level of social alignment for AI. Conversely, when these actors violate the social norms that they are subject to, they give rise to alignment problems that we may call social misalignment from violating existing norms. We start by discussing how informal and formal social norms address social alignment. Then we discuss why we believe that it is also increasingly desirable to impose new constraints directly on AI systems in addition to existing norms on their operators in order to guarantee that AI systems are socially aligned.

**Informal social norms** are constraints on agents' behavior that are enforced in an informal, decentralized way by a community. Social norms have evolved together with humanity to facilitate human cooperation. They can be seen in action, for example, when employees, consumers, or shareholders pressure companies to abstain from behaviors that they view as unethical.

The importance of informal social norms is frequently underemphasized—for example in economic analyses, when individuals are counterfactually depicted as perfectly selfish actors—and there is significant room for improving social AI alignment by establishing the right norms within the AI ecosystem (see e.g., Klinova, 2022). Social norms can be powerful in driving the behavior of individual humans. For example, social norms among AI developers as to what types of systems are considered ethical and desirable and what is considered unethical provide effective constraints on what systems AI companies develop. We can already see the effect of social norms on AI development in that many AI companies have begun creating codes of ethics or employing teams that directly focus on AI ethics and society (Bessen et al., 2021). Similarly, social norms among the broader public can translate into consumer pressure—not purchasing from AI companies that don't live up to their expectation of social alignment.

However, social norms alone are insufficient to govern our complex modern societies. They are most effective at the human community level. Non-human entities such as corporations and governments are not directly susceptible to social norms—only indirectly via their human agents—opening the door to what some have called "administrative evil" (see e.g. Young et al., 2021). More formal governance modes such as laws and regulations are therefore indispensable.

**Laws and Regulations** impose constraints on agents that are enforced with formal, state-backed or -administered penalties. There are several ways which such legal constraints can contribute to social alignment.

- Prohibitions and mandates: There are some uses of AI that society will deem too harmful to allow, and for these it may make sense to pass legislation which forbids them. For example, the Campaign to Stop Killer Robots has launched an effort to ban lethal autonomous weapons that could kill without human oversight.[19] Similarly, mandates can be enacted to ensure that AI systems meet certain socially desirable minimum standards, for example in the realm of safety. A closely related measure is to assign harmed individuals rights that can be enforced via litigation (see e.g., Kessler, 2010).

- Taxes and subsidies: Taxes and subsidies are a classic fix for externalities, and they apply just as well to the case of social AI alignment. They are preferable to outright bans and mandates when an activity creates externalities so the unregulated amount of that activity would be undesirable, but when a total ban would be excessive. Moreover, the revenue from taxes can be distributed to those who experience the harms.[20] For example, tracking of individuals represents a privacy intrusion but may also offer some useful benefits. Instead of banning it outright, imposing taxes or user fees may reduce it to a more desirable level. Similarly, if society's goals include an equitable income distribution, excessive automation that destroys jobs and undermines worker incomes could be taxed, with the resulting revenue distributed to workers losing their jobs, but it would be undesirable to ban automation.

## New norms for social AI alignment

The growing powers and capabilities of AI systems create new and ever more powerful ways in which the public interest may be infringed upon, i.e., new externalities, which call for new social norms and create new potential forms of social misalignment. When AI systems gain new capacities, society can be unprepared to govern. For example, society had one set of norms for surveillance and privacy when surveillance was very labor-intensive and correspondingly costly to undertake, so governments focused surveillance only on very high-value targets, such as suspected high-value criminals. Legal constraints on mass surveillance would have been redundant, given the high cost of surveillance. Now that AI systems can perform many forms of surveillance cheaply at large scale, new legal constraints on surveillance activities have become necessary.

More generally, every time a new AI capability is developed, they may bring up new social alignment problems that call for new social norms. Aside from privacy norms, additional examples in areas that we already touched upon include the need for new norms for AI systems that become increasingly adept at manipulating consumers, new fairness norms for AI systems that make high-impact decisions that have hitherto been reserved to humans. As the labor market effect of AI and other forms of automation become more severe—and more pernicious for workers—there is also a new need for norms for when and how AI developers should compensate the exposed workers. Even more starkly, if AI makes human workers economically redundant, society will need to establish new norms for how to provide humans with income when labor income is no longer an option (see Korinek and Juelfs, 2022).

---

[19] See https://www.stopkillerrobots.org/
[20] Allowing the harmed individuals themselves to impose a user fee is equivalent to taxing the harm and distributing the revenue to the harmed individuals.

**Social Alignment Norms Imposed on Whom?** Up until the recent past, governance to ensure the social alignment of AI systems has relied entirely on society imposing norms on the operators of AI systems, who were in turn charged with ensuring the alignment of their systems. This case, which we may call social alignment by extension, is illustrated in panel (a) of Figure 2, which employs arrows to indicate that an entity imposes norms on another entity. Such an arrangement would be all that is needed if (i) the operator was perfectly aligned with the social norms, and (ii) if the direct alignment between the operator and the AI system held perfectly.

However, when one of these two conditions is violated, it makes it desirable for society to directly impose social norms on AI systems, as illustrated in panel (b) of Figure 2. Let us consider each of the two conditions in turn.

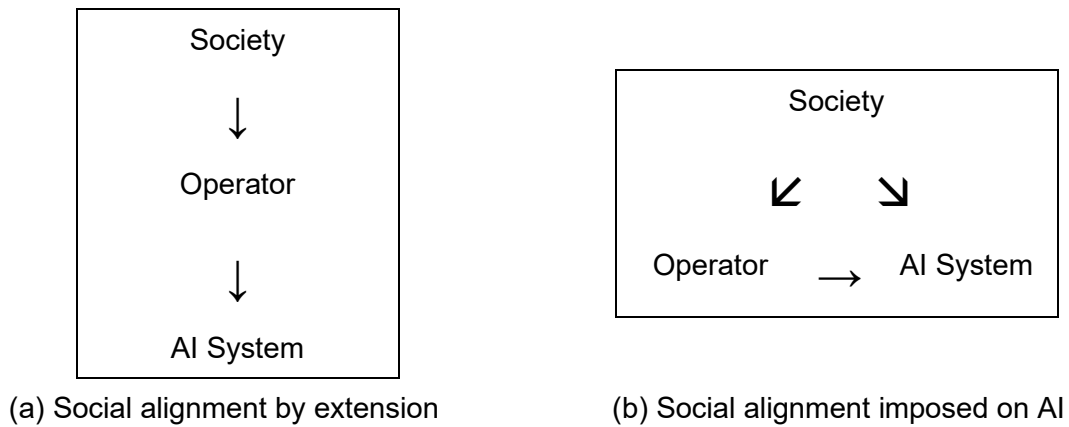

(a) Social alignment by extension  (b) Social alignment imposed on AI

Figure 2: Two modes of imposing social alignment norms on AI systems

When the operator of an AI system is not in compliance with social norms, then imposing social norms directly on AI systems may substitute for the operator's lack of compliance. Such an arrangement may also make it easier to monitor the social alignment of the operator. Consider for example an unethical corporation that pursues blind profit maximization to the detriment of other values of society. If the AI systems deployed by such a corporation need to satisfy certain enforceable norms—such as being unbiased—then the space for unethical behavior of the corporation is curtailed. In fact, norms imposed on AI systems may even make it possible to regulate behaviors that violate social norms but used to be difficult to regulate before. For example, when lending decisions were made individually by loan officers, it was harder to establish whether they were unbiased than it is with algorithms.

When an operator is generally aligned with social norms but has not fully solved the direct alignment problem between her and an AI system, then norms that are imposed directly on the AI system may also help. Such norms can be thought of as best practices, and they may contribute to all three steps of the direct alignment problem that we laid out above—determining the right goal, conveying the goal, and implementing the goal. For example, they may help a well-intended but inexperienced entrepreneur to ensure that the AI system she develops does not unintentionally impose harm on society.

As AI systems become more agentic and have ever more discretion over decisions that used to be reserved for humans, we believe that imposing norms directly on AI systems is becoming increasingly important.[21]

# V. Conclusion

As AI systems become more powerful and are deployed in a growing number of areas, aligning them with our goals becomes ever more vital. However, the expression "our goals" is often used too loosely. It is crucial to emphasize that AI alignment has two distinct dimensions—direct and social alignment. The two dimensions require somewhat different approaches, but we need to solve both to ensure a future that is desirable for humanity. Direct alignment ensures that AI systems pursue goals consistent with the objectives of their operators, irrespective of whether they impose externalities on other parties. By contrast, social alignment ensures that AI systems pursue goals that are consistent with the broader objectives of society, internalizing externalities and taking into account the welfare of everybody who is impacted by them.

Modern AI systems have the capacity to powerfully optimize for the goals that we endow them with. They are becoming better and better at doing what we are asking them to do and reaching their programmed goals—no matter if these goals correspond to our true goals or if we mistakenly assign them the wrong goals, for example excessively narrow subgoals that lead to disastrous unintended side effects because they fail to fully capture what we want.

Regarding direct alignment, we need to work on determining, conveying, and implementing the goals that we want AI systems to pursue in a robust manner. Regarding social alignment, society needs to determine what social goals and norms we want AI systems to pursue. Social preferences can only determine a partial ordering over all available choices. It is important to expand that ordering as much as possible by resolving social disagreements and conflicts, and to appeal to our better angels as we do this so that our preferences reflect our ethical values. To the extent that society finds agreement, it is also important to develop the right institutions to implement our preferences. We argue that this requires imposing norms on the developers and operators of AI systems as well as new norms that are directly imposed on AI systems.

As AI systems have become more powerful and their use in our world has become more widespread in recent years, we have also witnessed a growing number of cases of social alignment failures, from automated decision systems with biases against disadvantaged groups to social networks that increase polarization and undermine our political systems. Yet progress is continuing, and the powers of our AI systems are continuing to evolve. This makes it urgent to accelerate our efforts to better address the social alignment of AI. If we already have difficulty addressing the AI alignment problems we face now, how can we hope to do so in the future when the powers of our AI systems have advanced by another order of magnitude? Creating the right governance institutions to address the social AI alignment problem is therefore one of the most urgent challenges of our time.

---

[21] In related work, Korinek (2021) proposes the establishment of an AI Control Council to further these objectives.

# References

Armstrong, Stuart, Nick Bostrom, and Carl Shulman (2016), "Racing to the precipice: a model of artificial intelligence development." AI & Society 31(2): 201-206

Arrow, Kenneth J. (1950). "A Difficulty in the Concept of Social Welfare," Journal of Political Economy. 58 (4): 328–346.

Bajgar, Ondrej and Jan Horenovsky (2021), Human Rights as a Basis for Long-term AI Safety and Regulation.

Baum, Seth D (2020), Social Choice Ethics in Artificial Intelligence, AI & Society 35(1), pp. 165-176.

Bessen, James, Stephen M. Impink, Lydia Reichensperger and Robert Seamans (2021), "Ethics and AI Startups." https://scholarship.law.bu.edu/faculty_scholarship/1188

Bostrom, Nick (2014), *Superintelligence: Paths, Dangers, Strategies*, Oxford University Press.

Boukherouaa, El Bachir and Ghiath Shabsigh (2021), Powering the Digital Economy: Opportunities and Risks of Artificial Intelligence in Finance, Departmental Paper DP/2021/024, International Monetary Fund.

Christian, Brian (2020), *The Alignment Problem*, W.W. Norton.

Christiano, Paul (2018a), About AI alignment. AI Alignment. https://ai-alignment.com/about

Christiano, Paul (2018b), Clarifying "AI alignment," AI Alignment Forum. https://www.alignmentforum.org/posts/ZeE7EKHTFMBs8eMxn/clarifying-ai-alignment

Dafoe, Allan, Edward Hughes, Yoram Bachrach, Tantum Collins, Kevin R. McKee, Joel Z. Leibo, Kate Larson, Thore Graepel (2020), Open Problems in Cooperative AI, Technical Report, DeepMind.

Daniel C. Dennett (1987), *The Intentional Stance*, MIT Press.

Eckersley, Peter (2019), "Impossibility and Uncertainty Theorems in AI Value Alignment (or why your AGI should not have a utility function)," Proceedings of the AAAI Workshop on Artificial Intelligence Safety (SafeAI 2019), pp. 1-8.

Franklin, Stan and Art Graesser (1996), "Is it an Agent, or just a Program? A Taxonomy for Autonomous Agents," Proceedings of the Third International Workshop on Agent Theories, Architectures, and Languages, Springer. https://link.springer.com/chapter/10.1007/BFb0013570

Gabriel, Iason (2020), Artificial Intelligence, Values, and Alignment. Minds & Machines, 30, 411–437.
https://link.springer.com/content/pdf/10.1007/s11023-020-09539-2.pdf

Hubinger, Evan (2020), Clarifying inner alignment terminology, AI Alignment Forum.
https://www.alignmentforum.org/posts/SzecSPYxqRa5GCaSF/clarifying-inner-alignment-terminology

Jensen, Michael C.; Meckling, William H. (1976). "Theory of the firm: Managerial behavior, agency costs and ownership structure," Journal of Financial Economics 3(4): 305–360.

Juechems, Keno and Christopher Summerfield (2019), Where does value come from? Trends in Cognitive Sciences 23(10): 836-850.

Kessler, Daniel P. (2010), *Regulation vs. Litigation: Perspectives from Economics and Law*, University of Chicago Press.

Klinova, Katya (2022), "Governing AI to Advance Shared Prosperity," forthcoming in *Oxford Handbook of AI Governance*.

Korinek, Anton (2021), "Why we need a new agency to regulate advanced artificial intelligence: Lessons on AI control from the Facebook Files," Report, Brookings Institution, Dec. 8, 2021.
https://www.brookings.edu/research/why-we-need-a-new-agency-to-regulate-advanced-artificial-intelligence-lessons-on-ai-control-from-the-facebook-files/

Korinek, Anton and Megan Juelfs (2022), "Preparing for the (Non-Existent?) Future of Work," forthcoming in *Oxford Handbook of AI Governance*.

Ng, Andrew Y. and Stuart J. Russell (2000), "Algorithms for Inverse Reinforcement Learning," ICML '00: Proceedings of the Seventeenth International Conference on Machine Learning, pp. 663-670.

Ngo, Richard (2020), "AGI safety from first principles," AI Alignment Forum.
https://www.alignmentforum.org/s/mzgtmmTKKn5MuCzFJ

Orseau, Laurent, Simon McGregor McGill and Shane Legg (2018), "Agents and Devices: A Relative Definition of Agency." https://arxiv.org/abs/1805.12387

Russell, Stuart J. (2019), *Human Compatible: Artificial Intelligence and the Problem of Control*, Viking.

Russell, Stuart J. and Peter Norvig (2020), Artificial Intelligence: A Modern Approach, 4th US edition, Pearson.

Samuelson, Paul A. (1938). "A note on the pure theory of consumers' behaviour". Economica. New Series. 5(17): 61–71. doi:10.2307/2548836. JSTOR 2548836.

Weber, Max (1922), "Bureaucracy," translation of chapter 6 in *Wirtschaft und Gesellschaft*, Tübingen: Mohr.

Young, Matthew M., Johannes Himmelreich, Justin B. Bullock, and Kyoung-Cheol Kim (2021). "Artificial Intelligence and Administrative Evil." Perspectives on Public Management and Governance 4(3): 244–258. https://doi.org/10.1093/ppmgov/gvab006

Yudkowsky, Eliezer (2004), "Coherent Extrapolated Volition," Machine Intelligence Research Institute. https://intelligence.org/files/CEV.pdf